# Diffusion quantum Monte Carlo calculations of SrFeO$_3$ and LaFeO$_3$

Juan A. Santana,[1,2] Jaron T. Krogel,[1] Paul R. C. Kent,[3,4] and Fernando A. Reboredo[1, a]

[1] Materials Science and Technology Division, Oak Ridge National Laboratory, Oak Ridge, Tennessee 37831, USA

[2] Department of Chemistry, University of Puerto Rico at Cayey, P. O. Box 372230, Cayey, PR 00737-2230

[3] Center for Nanophase Materials Sciences, Oak Ridge National Laboratory, Oak Ridge, Tennessee 37831, USA

[4] Computer Science and Mathematics Division, Oak Ridge National Laboratory, Oak Ridge, Tennessee 37831, USA

**ABSTRACT:** The equations of state, formation energy and migration energy barrier of the oxygen vacancy in SrFeO$_3$ and LaFeO$_3$ were calculated with the diffusion quantum Monte Carlo (DMC) method. Calculations were also performed with various Density Functional Theory (DFT) approximations for comparison. DMC reproduces the measured cohesive energies of these materials with errors below 0.23(5) eV and the structural properties within 1% of the experimental values. The DMC formation energies of the oxygen vacancy in SrFeO$_3$ and LaFeO$_3$ under oxygen-rich conditions are 1.3(1) and 6.24(7) eV, respectively. Similar calculations with semi-local DFT approximations for LaFeO$_3$ yielded vacancy formation energies 1.5 eV lower. Comparison of charge density evaluated with DMC and DFT approximations shows that DFT tends to overdelocalize the electrons in defected SrFeO$_3$ and LaFeO$_3$. Calculations with DMC and LDA yield similar vacancy migration energy barriers, indicating that steric/electrostatic effects mainly determine migration barriers in these materials.

[a] Electronic mail: reboredofa@ornl.gov

This manuscript has been authored by UT-Battelle, LLC under Contract No. DE-AC05-00OR22725 with the U.S. Department of Energy. The United States Government retains and the publisher, by accepting the article for publication, acknowledges that the United States Government retains a non-exclusive, paid-up, irrevocable, worldwide license to publish or reproduce the published form of this manuscript, or allow others to do so, for United States Government purposes. The Department of Energy will provide public access to these results of federally sponsored research in accordance with the DOE Public Access Plan (http://energy.gov/downloads/doe-public-access-plan).

## I. INTRODUCTION

LaFeO$_3$ and SrFeO$_3$ are attractive multifunctional perovskite-type materials with practical applications in catalysis,[1–5] gas-sensing[6–8] and solid oxide fuel cells.[9–12] LaFeO$_3$ also exhibits multiple interesting physical properties, like multiferroic behavior,[13] colossal dielectric response,[13,14] and pressure-driven magnetic, structural, and electronic phase transitions.[15] Likewise, SrFeO$_3$ has unusual electronic, magnetic, structural and transport properties.[16,17] Moreover, fascinating physical properties have been reported for solid-solution and heterostructures[18] of LaFeO$_3$ and SrFeO$_3$.

Many of the physical properties and applications of these perovskite-type materials rely on the formation and mobility of oxygen vacancies ($V_O$). Accurate energetics of $V_O$ in these perovskite-type materials are, therefore, crucial for fundamental research as well as for the design and engineering of technological devices. For these reasons, a large number of computational calculations[17–29] have been performed to study $V_O$ in perovskites like LaFeO$_3$ and SrFeO$_3$. However, these calculations are based mainly on the standard Density Functional Theory (DFT) approximations, e.g., Local Density (LDA), semi-local General Gradient (GGA) and hybrid approximations. Such methods fail to describe insulating materials like LaFeO$_3$ (band gap of 2.1 – 2.3 eV),[30,31] in part, because of the residual self-interaction energy[32] inherent in the DFT approximations. For instance, LDA predicts LaFeO$_3$ to be metallic, GGA underestimates the band gap by over 1.3 eV,[23] and hybrid DFT approximations can overestimate the band gap by over 1.0 eV.[23] Introducing an on-site Hubbard model[33] correction (DFT+U) yields the experimental band gap of LaFeO$_3$,[23,31] and as a result, this method has been adopted to study $V_O$ in many perovskite-type materials.[18–26,29] However, a major drawback of DFT+U is that the parameter U depends on the chemical environment (e.g. oxidation state and surrounding ligands) of the atomic site where it is applied. Consequently, conventional DFT+U methods (using constant-U-values) are unreliable to study processes where the local environment of the atomic site changes.[34]

Because of the uncertainties in DFT approximations, it is necessary to validate DFT results. For many bulk properties (like structural and electronic properties), accurate and reliable experimental data is available and calculations can be easy benchmarked. In the case of ionic defects such as $V_O$, experimental measurements are complex and often indirect, and comparison with calculations is not always possible. The alternative route is then to benchmark the DFT results with many-body *ab-initio* calculations. Quantum Monte Carlo (QMC) methods, in particular the diffusion MC methods (DMC),[35] are the only computational techniques that have provided highly accurate many-body *ab-initio* simulations of ionic defects[36–46] in realistic materials at a manageable cost. Quantum Chemistry methods, while highly accurate, are prohibitively expensive to study ionic defects in materials.

In the present work, we have applied the DMC method to calculate the structural properties and the oxygen vacancy formation energy and migration energy barrier in the $SrFeO_3$ and $LaFeO_3$ perovskites. The aim of these calculations is twofold: *i*) assess the accuracy of our DMC methodology to study complex oxides and *ii*) provide a theoretical baseline for the oxygen vacancy formation and migration energy in $SrFeO_3$ and $LaFeO_3$. Our results show that DMC yields cohesive energy and structural properties in good agreement with the experimental values of these materials. The DMC oxygen vacancy formation energies in $SrFeO_3$ and $LaFeO_3$ under oxygen-rich conditions are 1.3(1) and 6.24(7) eV, respectively. We have carefully analyzed the possible sources of errors for the vacancy formation energies and estimated an overall accuracy of 0.2 eV. These DMC results can be employed to benchmark the DFT methodologies commonly used to study defects in complex oxides.

The paper is organized as follows. In Sec. II, we give a brief overview of the DMC method and computational details, including preliminary results and the possible sources of errors in our DMC calculations. In Sec. III, we start our discussion with the bulk properties of $SrFeO_3$ and $LaFeO_3$. We later discuss the DMC results for the energetics of the oxygen vacancy in these two materials in comparison with results from experiments and various DFT approximations. We conclude in Sec. IV with a summary of our calculations and findings.

## II. METHODOLOGY AND PRELIMINARY RESULTS

DMC is a stochastic method that projects out the ground state solution of a many-body problem by evolving a trial wavefunction $\Psi_T$ using the imaginary-time Schrödinger equation.[35] Various approximations, which can be systematically improved, are required for practical DMC calculations. The most significant approximations are the fixed-node (FN)[35] (or the generalized fixed-phase)[47] approximation and the use of pseudopotentials. Pseudopotentials are necessary because the computational cost of DMC scales proportional to $Z^{5.5-6.5}$, where $Z$ is the atomic number.[35]

The FN approximation is introduced to force the DMC solution of a many-fermion system to be antisymmetric.[35] This is accomplished by imposing the nodes of an antisymmetric trial wavefunction on the solution of the imaginary-time Schrödinger equation. If the trial wavefunction has the exact nodal surface, FN-DMC gives the ground state energy. Otherwise, the FN-DMC energy is an upper bound to the ground state energy (i.e., DMC is a variational method). A simple and straightforward approach to reduce fixed-node errors is to find a trial wavefunction with a nodal-surface that lowers the FN-DMC energy.[48] Other methods have been developed to improve the nodal surface during the random walk, and in turn, to increase the DMC accuracy (see Ref. 49 and references in there).

There are other approximations needed for practical DMC calculations, such as the short time-approximation and the use of supercells to simulate condensed matter, that also introduce errors,[35] i.e. time-step error, and one-body and two-body finite-size (FS) errors. In practice, however, these errors are easily tested and usually eliminated by extrapolation techniques. For FS errors, there are now various approaches to reduce or eliminate the need of extrapolation.[50–54] In what follows, we describe the DMC and DFT calculations and the pseudopotentials employed to calculated the equation of state (EOS) and the oxygen vacancy ($V_O$) formation energy in SrFeO$_3$ and LaFeO$_3$.

### A. DMC calculations and supercell models

DMC calculations were performed with QMCPACK[55] (http://qmcpack.org). Single-determinant Slater-Jastrow wavefunctions were used as guiding function. The Jastrow factor included one-, two- and tree-body terms with parameters optimized by variance minimization.[56] The Slater determinant was populated with single-particle orbitals generated with the plane-wave based code Quantum ESPRESSO.[57] The number of walkers in the DMC simulations was 2048 or more. Fixed-node, time-step and many-body finite-size[50–54] errors were analyzed and the results are discussed later in the paper.

The ionic cores were represented with norm-conserving pseudopotentials (NCPPs). We generated[58] the PPs for all relevant atoms. The O-, Sr-, La- and Fe-PPs are based on He-, Ni-, Pd- and Ne-core PPs, respectively. We will briefly describe the accuracy of our PPs below in Sec. C; further details can be found in our previous works.[59,60] The plane-wave energy cutoff was set to 4082 eV (300 Ry) because of the small-core PP of Fe. The scheme proposed by Casula[61] was used to treat the nonlocal part of the PPs within DMC and avoid numerical instabilities in the locality approximation.[62] The workflow automation system Nexus[63] was used to manage and monitor the various stages of the calculations.

We evaluated the EOS of cubic $SrFeO_3$ only for the ferromagnetic state (see discussion later in the paper). For $LaFeO_3$, the EOS was calculated for the ideal cubic structure and its known orthorhombic stable[64] structure, both with G-type antiferromagnetic ordering.[64] For the orthorhombic structure, the EOS was evaluated approximately by fixing the *a/b* and *c/b* ratios to the experimental values.[64]

The formation energy of $V_O$ in $SrFeO_3$ and $LaFeO_3$ under oxygen-rich conditions was calculated as $E^f[V_O] = E_{tot}[V_O] - E_{tot}[bulk] - \frac{1}{2}E_{tot}[O_2]$. $E_{tot}[V_O]$ and $E_{tot}[bulk]$ are, respectively, the total energies of the system with a defect and the equivalent bulk phase, and $E_{tot}[O_2]$ is the total energy of $O_2$. This approximation neglects thermal and vibrational effects on the defect formation enthalpy. However, these effects have been estimated to be below 0.1 eV for $V_O$ in $SrFeO_3$ and $LaFeO_3$.[23] The defect calculations were performed with a 2×2×2 (40 atom) supercell and twist boundary conditions on a 4×4×4 grid. (A DMC calculation with this model costs ~300000 compute hours on traditional processors, AMD Opteron 6274). For these calculations, the volumes of the pristine and defected supercells were kept fixed to the experimental values of $SrFeO_3$ (*a* = *b* = *c* = 3.851 Å)[16] and $LaFeO_3$ (*a* = 5.553, *b* = 5.563, and *c* = 7.862 Å).[64] The atomic positions of the pristine and defected systems were optimized within LDA+U until residual forces were below 0.02 eV/Å. The structures of the transition state for the migration of the $V_O$ in $LaFeO_3$ and $SrFeO_3$ were determined by applying the climbing-image nudged-elastic-band method[65] with 5 images.

### B. DFT calculations

Two types of DFT calculations were produced: *i*) DFT calculations on NCPP and *ii*) DFT calculations with standard methods and pseudopotentials used in the literature. The first group of DFT calculations includes LDA, LDA+U, GGA and GGA+U calculations and were performed to compare our DMC results for the bulk properties of SrFeO$_3$ and LaFeO$_3$ (results in Table 1 and Table 2). We include LDA/GGA+U results only for selected values of U, U = 3 and U = 6 eV. The second group of DFT calculations were carried out to compare the $E^f[V_O]$ calculated with DMC with various DFT aproximations: LDA, Perdew-Wang (GGA), +U[33] and Heyd-Scuseria-Ernzerhof (HSE)[66] hybrid functionals. LDA+U/GGA+U were performed with U values of 3, 5 and 7 eV, a range that enclosed values previously employed for Fe.[18,23,26] These calculations were performed with projector augmented wave (PAW)[67,68] ionic potentials as implemented in the Vienna Ab-initio Simulation Package (VASP).[69–71]

In general, DFT calculations with PAW potentials and our NCPPs yield similar results for $E^f[V_O]$. This indicates that the difference between DMC calculations and DFT are mainly due to a different treatment of many-body effect and not as much dependent on the pseudopotentials. For instance, $E^f[V_O]$ evaluated with LDA+U for SrFeO$_3$ and LaFeO$_3$ and our NCPPs are 1.63 and 5.30 eV, respectively. The corresponding values evaluated with PAW potentials are 1.67 eV and 5.10 for SrFeO$_3$ and LaFeO$_3$, respectively. However, calculations are less computationally expensive with PAW potentials because the required wavefunction energy cutoff is only 550 eV. For the DFT calculations with VASP, we used O, Sr and La PAW potentials with 6, 10 and 11 valence electrons, respectively. For Fe, we performed calculations with various of the PAW potentials available in VASP, specifically the potentials with 8 (PAW-Fe), 14 (PAW-Fe_pv) and 16 (PAW-Fe_sv) valence electrons. Our results showed that $E^f[V_O]$ evaluate with DFT+U and Fe_pv could be up to 0.4 eV higher than values from calculations with PAW-Fe and PAW-Fe_sv. Calculations with PAW-Fe and PAW-Fe_sv yield $E^f[V_O]$ within 0.1 eV. Therefore, the DFT results for $E^f[V_O]$ that are discussed in this work are based on calculations with the PAW-Fe potential.

### C. Possible sources of errors and preliminary results

There are various sources of errors in the simulation of materials with ionic defects.[72] In the present cases, these errors can come from *i*) the approximations in DMC (e.g., pseudopotentials, fixed-node, time-step and many-body FS effects) and *ii*) the modeling of defects with the supercell approximation (e.g., overlapping of defect level, electrostatic and elastic interactions).[72] In what follows, we discuss our efforts to estimate these errors.

*Pseudopotentials:* Reliable pseudopotentials are key in DMC calculations. In practice, using PPs with cores as small as possible reduces these errors. We used relatively small-core PPs for Sr and La, and we have tested them within DMC by evaluating the ionization potentials (IP) of each atom and selected bulk properties for the binary oxides SrO and $La_2O_3$.[59] We also tested the Fe-PP by evaluating the IP[60] of Fe and the cohesive energy of FeO (see the Supplemental Material section for details). The DMC results for the IPs of Sr, La and Fe deviate from the experimental values by an average of 0.15 eV. For the SrO, $La_2O_3$ and FeO oxides, our DMC-PPs calculations reproduce the measured cohesive energies with errors below 0.15 eV. Similar, our DMC-PPs calculations yield highly accurate lattice constants for SrO and $La_2O_3$, within 0.5% of the experimental values; see Refs. 59 and 60 for further details of our PPs.

*Fixed-node errors:* Single-particle orbitals generated with a Hubbard-corrected functional (LDA+U) instead of LDA can be used to explore and reduce the fixed-node errors (see the Supplemental Material section for details). Single-particle orbitals generated with U = 3.0 and 6.0 eV yielded the lowest DMC energy for $SrFeO_3$ and $LaFeO_3$, respectively. The total DMC energy per formula unit (f.u.) of $SrFeO_3$ and $LaFeO_3$ are 0.26(2) and 0.34(1) eV/f.u lower when single-particle orbitals are generated using these U-values instead of LDA. Single-particle orbitals generated with these U-values also yield the lowest DMC energy for defected (with $V_O$) $SrFeO_3$ and $LaFeO_3$. In these cases, the total DMC energy is 0.43(2) and 2.09(2) eV lower if single-particle orbitals are generated using these U-values instead of LDA. The fixed-node errors are different for pristine and defected systems. Therefore, these errors cannot be expected to cancel out for $E^f[V_O]$ in these systems.

*Time-step errors:* To estimate time-step errors, we performed calculations with three time-steps (0.02, 0.01, and 0.005 $Ha^{-1}$). These calculations were performed with a $\sqrt{2}\times\sqrt{2}\times2$ (20 atom) supercell and twist boundary conditions on a 6×6×6 grid (see the Supplemental Material section for details). From these calculations, we extrapolated the calculated properties to 0 time-step. For $LaFeO_3$, the effect of time-step is very small. For instance, the formation energy of $V_O$ evaluated with a time-step of 0.01 $Ha^{-1}$ is only 0.06(2) eV lower than the value extrapolated to 0 time-step. For SrFeO$_3$, the formation energy evaluated with a time-step of 0.01 Ha$^{-1}$ is, on the other hand, significantly lower than the value extrapolated to 0 time-step, i.e. 0.87(8) eV. Calculations with the 40-atom supercell model were performed with a time-step of 0.01 Ha$^{-1}$ and corrected by the estimated time-step errors to extrapolate to 0 time-step.

*Many-body finite-size errors:* The use of supercells to simulate condensed matter introduces FS errors in all QMC calculations.[50] These errors are common to most many-body methods and can be divided into one and two-body FS errors. One-body errors come from an artificial momentum quantization due to periodic boundary conditions imposed to the electrons in the simulation cell.[50] The two-body error arises from the artificial periodicity of the exchange-correlation hole.[50] In the present calculations, one-body FS errors have been treated employing twisted boundary conditions.[52] At the DFT level, the supercell model and k-point grids that we employed yield total energies for defected systems that are converged within 1 meV. Based on these results, we expected one-body FS errors in DMC to be small, likely below the statistical error of our DMC results. Two-body FS errors are usually eliminated by extrapolation techniques. Such calculations are impractical for the present systems. However, we have evaluated two-body FS errors approximately with the method proposed by Kwee, Zhang and Krakauer[54] and the model periodic Coulomb (MPC) interaction[50,53] corrected for kinetic contributions[50,53] (see the Supplemental Material section for details). For the formation energy of $V_O$ in SrFeO$_3$ and LaFeO$_3$, two-body FS errors are only 0.1(1) eV because of error cancellation.

*Supercell finite-size errors:* Previous DFT calculations[73] have shown that a supercell model of 40 atoms yields formation energies of $V_O$ in LaFeO$_3$ that are converged with respect to defect concentration. In our test calculations, we found that the formation energy of $V_O$ in LaFeO$_3$ is similar within 0.1 eV when evaluated with GGA+U (5 eV) in 40 and 160 atom supercells. For SrFeO$_3$, GGA+U (U = 5 eV) calculations with a 160-atom supercell yield a formation energy of $V_O$ 0.2 eV lower than with the 40-atom model. Increasing the supercell to a 320-atom model further reduces the formation energy only by 0.01 eV. These finite-size effects for SrFeO$_3$ come, in part, from elastic interactions.

In fact, allowing volume relaxation in the 40-atom supercell model of SrFeO$_3$ and LaFeO$_3$ lowers the formation energy of $V_O$ by 0.07 and 0.03 eV, respectively.

Finally, we mention the used of DFT-optimized atomic positions in the DMC calculations. Ideally, atomic positions should be optimized with DMC as well, but DMC calculation of interatomic forces in large systems is impractical now. For energy differences (e.g., defect formation energy), taken the atomic positions from a DFT approximation introduces only minor errors.[43,74]

## III. RESULTS AND DISCUSSION

### A. Equations of state of SrFeO$_3$ and LaFeO$_3$

LaFeO$_3$ is an insulator with a band gap of 2.1 to 2.3 eV.[30,31] It has an orthorhombic structure with a $a^-a^-c^+$ octahedral tilt pattern[75] and G-type antiferromagnetic ordering.[64,76] On the other hand, SrFeO$_3$ is a metal with a cubic structure and a helical spin order below the Néel temperature of 133(1) K.[77,78] The incommensurate helical order indicates a delicate competition between ferromagnetic and antiferromagnetic interactions on the threshold of a metal-insulator transition.[77–79] Previous LDA+U (U = 5.4 eV) calculations[80] showed that the collinear magnetic order with the lowest energy in SrFeO$_3$ is the ferromagnetic state. Based on these calculations,[80] the second most stable collinear magnetic state is the antiferromagnetic ordering between adjacent planes of Fe along the z-axis (A-type), which is 0.072 eV/f.u. higher in energy that the ferromagnetic state. We found similar stability order with our GGA and GGA+U calculations; GGA and GGA+U (U = 3.0 eV) yield energy separations between these two magnetic states of 0.071 eV/f.u. and 0.067 eV/f.u., respectively. Based on these results, we evaluated the EOS of SrFeO$_3$ with DMC only for the ferromagnetic state. We do not expect DMC to yield a different relative order for these collinear magnetic states. However, the magnitude of the energy separation could be different in DMC.[81,82]

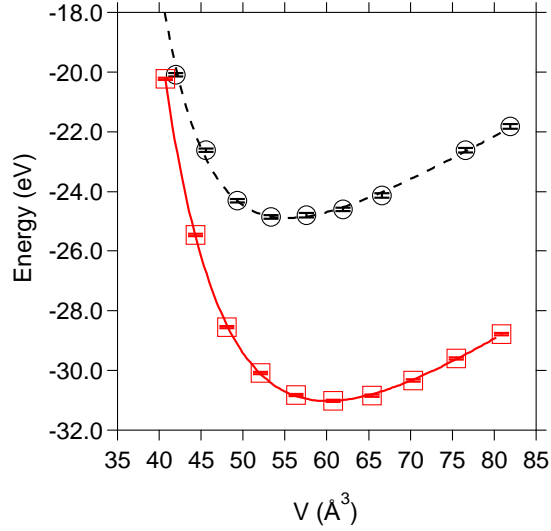

**Figure 1.** DMC energies versus volume per formula unit of $SrFeO_3$ (circles) and $LaFeO_3$ (squares) together with fitted equations of state (Murnaghan). Energies are per formula unit and relative to the Sr, La, Fe and O atoms. The statistical uncertainty in DMC is smaller than the symbol size.

The EOS of $SrFeO_3$ and $LaFeO_3$ evaluated with DMC are shown in Figure 1. The energies were fitted to Murnaghan EOS, and the derived structural parameters are included in Table 1 and Table 2. Results from LDA, GGA and +U functionals are also included in the tables for comparison. DMC closely reproduces the measured lattice constants of both $SrFeO_3$ and $LarFeO_3$; the error is at the 1% level. The various DFT approximations also perform relatively well for the lattices constants. For instance, LDA and LDA+U show some deviation (1.6%) for $SrFeO_3$, but agree with experiment at the 1% level for $LaFeO_3$. Similarly, GGA and GGA+U yield lattice constants in agreement with experiment for $SrFeO_3$ and $LaFeO_3$.

**Table 1.** Lattice constant $a$ (Å), bulk modulus B (GPa) and its first derivative B′, and cohesive energy (eV) of $SrFeO_3$ evaluated with DMC, LDA, GGA and +U (U = 3.0 eV) functionals. Available experimental results are included for comparison. The statistical uncertainty is provided in parenthesis.

| Method | $a$ | B | B′ | Energy |
|---|---|---|---|---|
| DMC | 3.809(1) | 199.4(2) | 7.84(1) | 24.88(5) |
| LDA | 3.743 | 174.12 | 4.67 | 31.57 |
| LDA+U | 3.786 | 158.49 | 4.34 | 29.92 |
| GGA | 3.840 | 140.20 | 4.22 | 24.97 |
| GGA+U | 3.888 | 132.24 | 4.00 | 23.66 |
| Expt. | 3.851[a] | | | 24.89[b] |

[a] Ref. 16; [b] Cohesive energy evaluated with the experimentally estimated enthalpy of formation of $SrFeO_3$ (Ref. 83) and the cohesive energies of SrO and $Fe_2O_3$ (Refs. 84).

The various DFT approximations yield rather similar values for the bulk modulus (B) and its pressure derivative (B′) for $SrFeO_3$ (Table 1) and $LaFeO_3$ (Table 2). The corresponding values from DMC are slightly larger, particularly B′. We did not find measurements to compare with our calculations. Previous DFT calculations with the full-potential linear muffin-tin orbital method[85] and a hybrid full potential linear augmented plane wave + local orbital methods[86] also yielded large values of B for $SrFeO_3$, 205.24 and 196 GPa, respectively.

**Table 2.** Lattice constants *a, b* and *c* (Å), bulk modulus B (GPa) and its first derivative B′, and cohesive energy (eV) of cubic and orthorhombic $LaFeO_3$ evaluated with DMC, LDA, GGA and +U (U = 6.0 eV) functionals. Available experimental results are included for comparison. The statistical uncertainty is provided in parenthesis.

| Method | System | a, b, c | B | B′ | Energy |
|---|---|---|---|---|---|
| DMC | Cubic $LaFeO_3$ | 3.906(1) | 207.3(1) | 5.52(1) | 30.76(5) |
| LDA+U | | 3.900 | 189.3 | 4.05 | 34.07 |
| GGA+U | | 3.970 | 167.4 | 3.88 | 27.47 |
| DMC | Orthorhombic $LaFeO_3$ | 5.496(1), 5.506(1), 7.781(1) | 189.4(1) | 6.19(1) | 31.01(5) |
| LDA+U | | 5.487, 5.497, 7.768 | 170.8 | 4.26 | 34.37 |
| GGA+U | | 5.831, 5.842, 8.256 | 141.69 | 4.22 | 27.871 |
| Expt. | | 5.553, 5.563, 7.862[a] | | | 30.78[b] |

[a] Ref. 64; [b] Cohesive energy evaluated with the experimentally estimated enthalpy of formation of $LaFeO_3$ (Ref. 83 and 87) and the cohesive energies of $La_2O_3$ (Ref. 88 and 89) and $Fe_2O_3$ (Refs. 84).

The cohesive energy of SrFeO3 and LaFeO3 evaluated with DMC agrees well with experimental values (Table 2). The deviations are only 0.01(5) and 0.24(5) eV/f.u. for SrFeO3 and LaFeO3, respectively. A similar good agreement between DMC and measured cohesive energies was reported for the complex oxide BaTiO3.[90] As usual, LDA and LDA+U overestimate the cohesive energies of SrFeO3 and LaFeO3 (by over 3 eV/f.u.). GGA and GGA+U yield cohesive energies close to experiments for SrFeO3, but deviate by over 2 eV/f.u. for LaFeO3. The cohesive energy of LaFeO3 was evaluated for the cubic and orthorhombic structures. The energy difference between these two structures quantifies the structural change and the octahedral tilt present in orthorhombic structures. The DMC value is 0.25(7) eV/f.u, which is close to 0.30 and 0.40 eV/f.u. yielded by LDA+U and GGA+U, respectively.

DMC clearly outperforms the standard DFT approximations in describing the bulk cohesive energies of LaFeO3 and SrFeO3. For structural properties, like the lattice constants, both methods perform reasonably well. These results are encouraging because one can expect DMC to provide a better description of the energetics of defects in these materials. The results also provide further support for the use of DFT-optimized atomic positions in the DMC calculations of defects as the structural properties of bulk materials obtained with the various methods are not too different.

### B. Formation energy of $V_O$ in SrFeO3 and LaFeO3 from DMC

The DMC formation energy for the oxygen vacancy ($E^f[V_O]$) in SrFeO3 and LaFeO3 are 1.3(1) and 6.24(5) eV, respectively, under oxygen-rich conditions (see Sec. II.A). The calculated $-E^f[V_O]$ for SrFeO3 can be compared with oxidation enthalpies derived from thermogravimetric studies[83,91,92] of La$_{1-x}$Sr$_x$FeO$_{3-\delta}$. For instance, the enthalpy of oxidation for $1.0 < x > 0.5$ was reported[83] to be 0.73(16) eV. A similar value of 0.83 eV has been reported[92] for SrFeO$_{3-\delta}$. For $0.5 < x > 0.1$, a value of 1.04(26) eV was reported,[83] which is similar to 1.09(13) eV reported earlier[91] for $0.6 < x > 0.1$. In the cases where $1.0 < x > 0.1$, the oxidation enthalpy derived from experiments correspond to the oxidation reaction $V_O + Fe_{Fe}^{3+} + 0.5O_2 \rightarrow O_O^x + 2Fe_{Fe}^{4+}$. For LaFeO3, the corresponding oxidation enthalpy is derived from experiments by extrapolating to $x = 0$. The oxidation enthalpy in this case must correspond to the oxidation reaction $V_O + Fe_{Fe}^{2+} + 0.5O_2 \rightarrow O_O^x + 2Fe_{Fe}^{3+}$.

The change in oxidation state can be accounted for employing the measured enthalpy of ionization $2Fe_{Fe}^{3+} \rightarrow Fe_{Fe}^{2+} + Fe_{Fe}^{4+}$ at $x = 0$. Employing results reported in Ref. 91 for the oxidation enthalpy and enthalpy of ionization at $x = 0$, a value of 5.44(1.04) eV is obtained[93] for the oxidation enthalpy in LaFeO$_3$. This value compared with our DMC results for $V_O$ in LaFeO$_3$.

The large difference in the formation energy of $V_O$ in SrFeO$_3$ and LaFeO$_3$ was previously rationalized based on DFT calculations and analyses.[21,23] According to the previous results, the higher formation energy for $V_O$ in LaFeO$_3$ comes from the reduction of Fe ions from Fe$^{3+}$ to Fe$^{2+}$ upon formation of $V_O$, which introduces additional electron-electron repulsions.[21] In the case of SrFeO$_3$, the formation of $V_O$ compensates holes in the oxygen sublattice,[23] resulting in a low defect formation energy. To corroborate these models, we determined the location of the impurity electrons in defected SrFeO$_3$ and LaFeO$_3$. To probe the location of the impurity electrons upon formation of $V_O$, we calculated the difference in electron density ($\Delta\rho$) between the neutral and vertically (unrelaxed) ionized vacancy. We evaluated $\Delta\rho$ with a DMC estimator as well as with a combination of variational Monte Carlo (VMC) and DMC estimators (extrapolated estimator).[35] No quantitative differences were found between $\Delta\rho$ evaluated with the two estimators. Results from DMC for SrFeO$_3$ and LaFeO$_3$ are shown in Figure 2. Indeed, $\Delta\rho$ indicates that upon formation of $V_O$ the impurity electrons are localized on different sites in SrFeO$_3$ and LaFeO$_3$. The impurity electrons are mainly localized on the O sites in defected SrFeO$_3$ and on the Fe sites in the case of defected LaFeO$_3$.

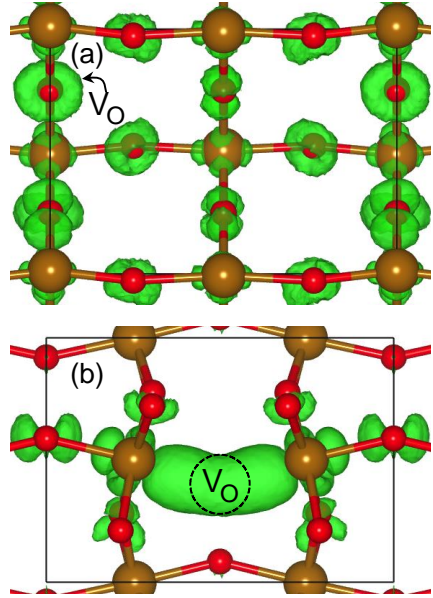

**Figure 2**. Difference in electron density (Δρ) between neutral and vertically (unrelaxed) ionized $V_O$ in (a) SrFeO$_3$ and (b) LaFeO$_3$ evaluated with DMC. Small and large spheres correspond to O and Fe atoms, respectively. Sr and La atoms are not displayed for the sake of clarity. Images were generated with VESTA[94] with an isosurface level of 0.005 $e^-/Bohr^3$.

## C. Comparison of $E^f[V_O]$ evaluated with DMC and DFT approximations

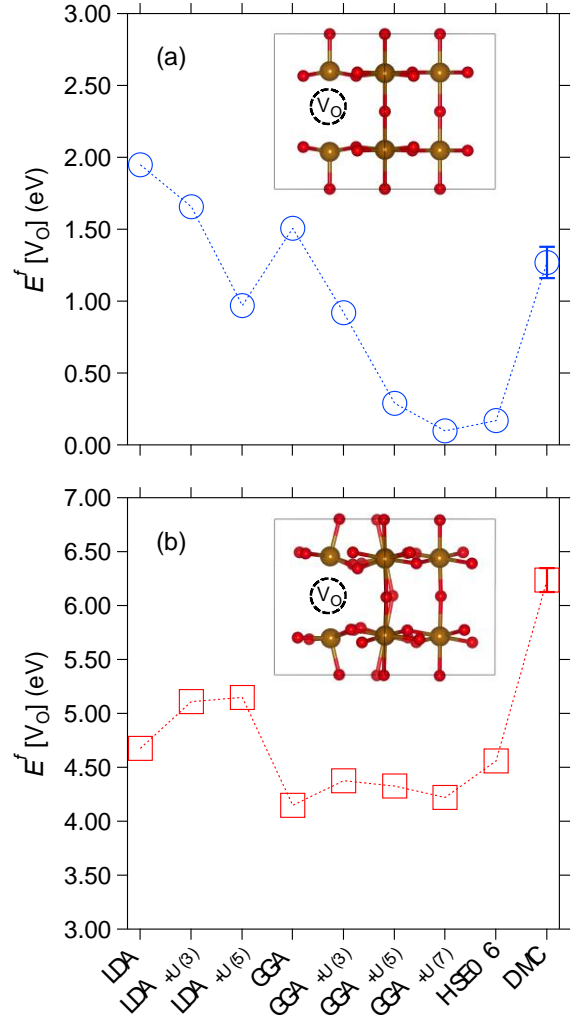

**Figure 3.** Formation energy of the oxygen vacancy under O-rich conditions in (a) SrFeO$_3$ and (b) LaFeO$_3$ evaluated with DMC and various DFT approximations. The statistical uncertainty in DMC for LaFeO$_3$ is smaller than the symbol size. Inset: 40 atom supercells of defected SrFeO$_3$ and LaFeO$_3$ (Sr and La atoms are not displayed for clarity).

$E^f[V_O]$ evaluated with DMC and various DFT approximations are shown in Figure 3 for SrFeO$_3$ and LaFeO$_3$. None of the DFT approximations consistently reproduce the DMC results for the two materials. The best agreement is found in the case of SrFeO$_3$, where LDA+U (U = 5 eV) and GGA+U (U = 3 eV) yield $E^f[V_O]$ within 0.3 eV of the DMC value. In LaFeO$_3$, LDA+U and GGA+U yield $E^f[V_O]$ lower than DMC by 0.5 and 1.5 eV, respectively. HSE performs poorly for both materials, yielding $E^f[V_O]$ much lower than DMC (> 1 eV). Figure 3 also shows that $E^f[V_O]$ has a different sensitivity to the U parameter in the two materials. $E^f[V_O]$ decreases monotonically for increasing U-values in the case of SrFeO$_3$. On the other hand, $E^f[V_O]$ increases with increasing U in LaFeO$_3$, reaching a plateau region for U values of about 5 eV. The sensitivity of $E^f[V_O]$ to the U-value was previously pointed out in Refs. 19, 23 and 25.

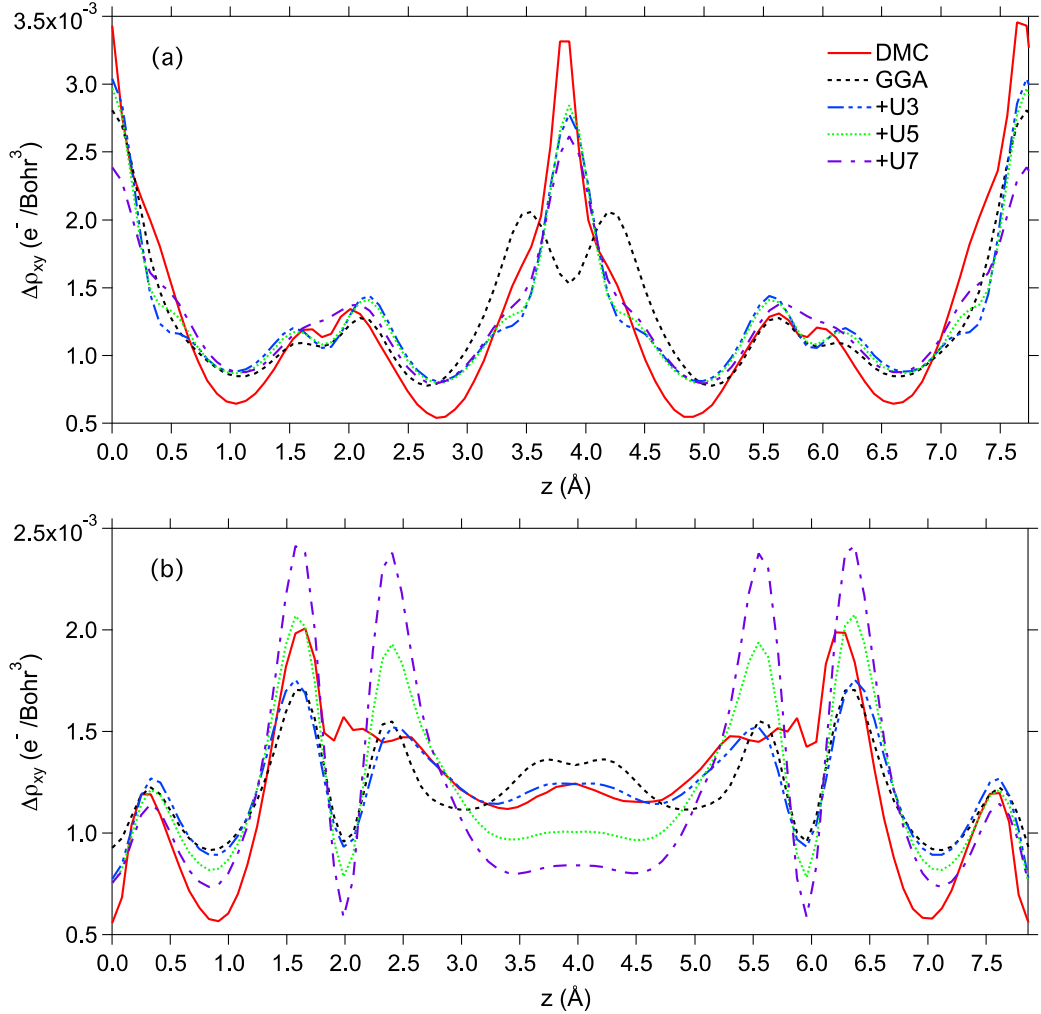

**Figure 4**. Planar average of the difference in electron density ($\Delta\rho_{xy}$) between neutral and vertically ionized $V_O$ in (a) SrFeO$_3$ and (b) LaFeO$_3$. Calculations were performed with DMC, GGA and GGA+U (U = 3, 5 and 7). $V_O$ is located at the origin ($z \approx 0$ Å) and center ($z \approx 4$ Å) of the SrFeO$_3$ and LaFeO$_3$ supercells, respectively; see Figure 2 for reference.

For a qualitative assessment of the differences between $E^f[V_O]$ evaluated with DMC and GGA, we compared the difference in electron density (Δρ) between the neutral and vertically ionized $V_O$ evaluated with DMC, GGA and GGA+U. Δρ probes the location of the impurity electrons upon formation of $V_O$ (Figure 2). We evaluated the *xy*-planar average of Δρ to perform the comparison. Results for the *xy*-planar average of Δρ ($\Delta\rho_{xy}$) are shown in Figure 4a and Figure 4b for SrFeO$_3$ and LaFeO$_3$, respectively. For SrFeO$_3$, DMC and GGA yield a different profile for $\Delta\rho_{xy}$ in the center of the supercell. The difference arises because in GGA the impurity electrons are mainly located in the Fe site; in DMC (Figure 2), they are in the oxygen sites. Introducing the U parameter corrects this error in GGA, and GGA+U resemble better the DMC $\Delta\rho_{xy}$ in SrFeO$_3$. For LaFeO$_3$, the difference between DMC, GGA, and GGA+U are more significant than in SrFeO$_3$. Applying a U-value of 3 eV yields $\Delta\rho_{xy}$ closer to DMC in some regions in LaFeO$_3$ (e.g., $z \approx 3$ Å). However, GGA+U with this U-value fails to reproduce the DMC results in other regions (e.g., $z \approx 1$, 1.5 and 2 Å). Increasing the U-value to 5 eV improves the description around $z \approx 1.5$ Å, but at the expense of a poorer description in other regions ($z \approx 2.5$ and 4 Å). It is clear from Figure 3 and Figure 4 that as $\Delta\rho_{xy}$ evaluated with GGA+U gets closer to the DMC result, $E^f[V_O]$ evaluated with DMC and GGA+U are also more alike. However, GGA and GGA+U consistently overestimate $\Delta\rho_{xy}$ in many regions (e.g., $z \approx 1$ Å), and $E^f[V_O]$ evaluated with DMC is not fully reproduced with GGA+U. We note that the challenge of improving DFT-based approaches to correctly reproduce the electron density applies not only to the solid state but also to molecular transition metal complexes.[95]

### D. Energy barrier for the migration of V$_O$ in SrFeO$_3$ and LaFeO$_3$

The migration energy barriers evaluated with DMC, GGA, GGA+U and LDA+U are included in Table 3. In general, the GGA, GGA+U and LDA+U calculations with PAW potentials yield values in the 0.6 – 0.9 eV range for both materials, which agree well with previous DFT calculations: 0.75[22] and 0.79[23] eV for LaFeO3 and 0.76[22], 0.78[96] and 0.89[97] eV for SrFeO$_3$. These energy barriers also follow in the range of available measurements: 0.77 eV[98] and 1.10 eV[99] for LaFeO$_3$ and 0.9 eV[100] and 0.82 eV[101] for SrFeO$_3$. The similarity of the energy barriers indicates that steric/electrostatic effects mainly determine migration barrier in SrFeO$_3$ and LaFeO$_3$. The DMC energy barriers are 0.48(2) and 1.06(5) eV in SrFeO$_3$ and LaFeO$_3$, respectively. The DMC migration energy barriers are calculated with the atomic structure of the migration transition state evaluated with LDA+U and PAW-Fe potential. The DMC results deviate by ~0.3 eV from the DFT calculations based on PAW potentials. The deviation is likely coming from using the atomic structure evaluated with PAW potentials. This is partially demonstrated by the fact that calculations with LDA+U and NCPPs on the atomic structure used in DMC yield energy barriers close to DMC. The implication here is that the atomic structure of the migration transition state is sensitive to the employed pseudopotentials.

**Table 3.** Energy barrier (eV) for the migration of oxygen vacancies in SrFeO$_3$ and LaFeO$_3$ calculated with DMC and DFT approximations. The statistical uncertainty is provided in parentheses.[a]

| Method | SrFeO$_3$ | LaFeO$_3$ | |
|---|---|---|---|
| GGA | 0.76 | 0.58 | PAW-Fe |
| GGA+U(3) | 0.69 | 0.82 | PAW-Fe |
| GGA+U(5) | 0.75 | 0.83 | PAW-Fe |
| GGA+U(7) | 0.81 | 0.88 | PAW-Fe |
| LDA+U[a] | 0.58 | 0.69 | PAW-Fe |
| LDA+U[a,b] | 0.40 | 0.95 | NCPP |
| DMC[b] | 0.48(2) | 1.06(5) | NCPP |

[a] Calculations with LDA+U with U= 3 for SrFeO$_3$ and U = 6 for LaFeO$_3$. [b] The atomic structures of the migration transition state are taken from calculations with LDA+U and the PAW-Fe potential.

### E. Comparison with previous DMC defect formation energy

Previous studies have shown significant differences between defect formation energies evaluated with DMC and DFT approximations.[43,36,41,42] For instance, the formation energies of Al[41] and Si[36] self-interstitial were found 0.3 and 1 eV higher when evaluated with DMC instead of GGA. Similarly, $E^f[V_O]$ in MgO[42] was found to be approximately 0.5 eV higher in DMC than GGA. In the case of $V_O$ in ZnO, we recently reported[43] that DMC yields $E^f[V_O]$ 0.5 – 1.5 eV higher than GGA and the HSE06 hybrid functional. The error for $E^f[V_O]$ in MgO has been associated[42] with the overdelocalization in GGA. The overdelocalization arises from the residual spurious self-interaction energy[32] inherent in LDA/GGA. This is also responsible for the underestimation of the band gap of insulators. $E^f[V_O]$ evaluated with GGA is low for insulators because the occupied deep level of $V_O$ is too close to the VBM.

There is convincing evidence supporting that the significant differences between DMC and DFT approximations for defect formation energies are, indeed, primarily due to the self-interaction errors in DFT. First, self-interaction errors are small for orbitals delocalized over extended systems.[32] Therefore, GGA should yield defect formation energy in closer agreement with DMC for metallic systems. Indeed, this is the case for the formation energies of Al[41] self-interstitial and $V_O$ in SrFeO$_3$ (Figure 3), where GGA and DMC results are with 0.4 eV. Second, methods yielding an improved description of the band gap should also yield $E^f[V_O]$ closer to DMC. In our previous work,[43] we showed that this is indeed the case for ZnO. For instance, HSE with a 0.38 fraction of nonlocal Fock-exchange (HSE38) yields the band gap of ZnO in agreement with experiment and $E^f[V_O]$ 0.5 closer to DMC than GGA. Similar results are also found in the present work for LaFeO$_3$. The use of GGA+U describes the band gap of LaFeO$_3$ better[23,31] than GGA, and, as shown in Figure 3, it also yields $E^f[V_O]$ closer to DMC.

The picture that emerges from the results described above for insulators is that providing a better description for the band gap within DFT does improve the description of defect energetics. However, formation energies evaluated with these improved DFT methods can still be significantly lower than in DMC. For instance, $E^f[V_O]$ evaluated with GGA+U is over 1.5 eV lower than DMC (Figure 3). In the case of ZnO, $E^f[V_O]$ evaluated with HSE38 is over 0.5 eV lower than the DMC value.

## IV. SUMMARY

In summary, we have applied the diffusion quantum Monte Carlo method to study the complex oxides SrFeO$_3$ and LaFeO$_3$. We evaluated the equation of state, the oxygen vacancy formation and migration energies in these materials. The DMC errors are below 0.23(5) eV for the cohesive energy and at the 1% level for the structural properties of SrFeO$_3$ and LaFeO$_3$. DMC yields oxygen vacancy formation energies of 1.3(1) and 6.24(7) eV for SrFeO$_3$ and LaFeO$_3$, respectively, under oxygen-rich conditions. Taken in consideration the possible sources of errors for these formation energies, we estimated an overall accuracy of 0.2 eV. We compared these DMC formation energies with some of the DFT approximations (LDA, GGA, +U and hybrid functionals) commonly used to study defects in complex oxides. The comparison shows that GGA+U can reproduce the DMC oxygen vacancy formation energy for a metallic system like $SrFeO_3$, but severely underestimates it for an insulator like $LaFeO_3$. Charge density analyses indicate that the discrepancy between the two methods is rooted on the fact that the DFT approximation over-delocalizes impurity electrons.

## SUPPLEMENTARY MATERIAL

See the Supplementary Material for details of the DMC calculations of FeO and the fixed-node, time step and finite-size errors in the DMC calculations of $SrFeO_3$ and $LaFeO_3$.

## ACKNOWLEDGMENT

We thank Rohan Mishra, Albina Y. Borisevich and Sokrates T. Pantelides for helpful discussions. The work was supported by the Materials Sciences & Engineering Division of the Office of Basic Energy Sciences, U. S. Department of Energy. Paul R. C. Kent was supported by the Scientific User Facilities Division, Office of Basic Energy Sciences, U. S. Department of Energy. Computational resources were provided by the Oak Ridge Leadership Computing Facility at the Oak Ridge National Laboratory, supported by the Office of Science of the U. S. Department of Energy under DE-AC05-00OR22725.

# Supplemental Material for

# Diffusion Quantum Monte Carlo calculations of SrFeO$_3$ and LaFeO$_3$

Juan A. Santana,[1,2] Jaron T. Krogel,[1] Paul R. C. Kent,[3,4] and Fernando A. Reboredo[1, a]

[1] Materials Science and Technology Division, Oak Ridge National Laboratory, Oak Ridge, Tennessee 37831, USA

[2] Department of Chemistry, University of Puerto Rico at Cayey, P. O. Box 372230, Cayey, PR 00737-2230

[3] Center for Nanophase Materials Sciences, Oak Ridge National Laboratory, Oak Ridge, Tennessee 37831, USA

[4] Computer Science and Mathematics Division, Oak Ridge National Laboratory, Oak Ridge, Tennessee 37831, USA

## Contents

[a] Electronic mail: reboredofa@ornl.gov

# I. SELECTED DMC CALCULATIONS OF FEO

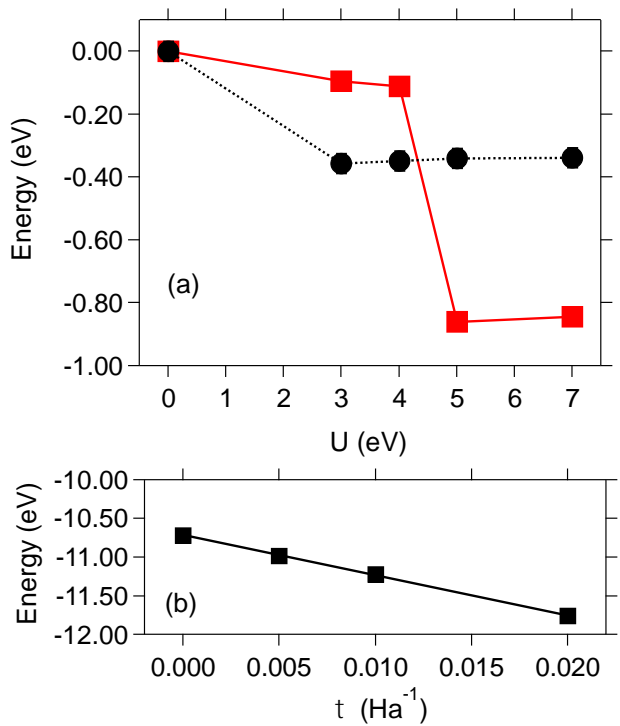

**Figure S1**. (a) DMC energy evaluated with single-particle orbitals generated with LDA+U. Results are included for the Fe atom (circles) and the FeO crystal (squares). Energies are given per formula unit and relative to DMC results with orbitals from LDA. (b) DMC energy as a function of imaginary time-step for FeO; extrapolated energies are included at time-step 0. Energies are given per formula unit and relative to the Fe and O atoms. The statistical uncertainty in DMC results is smaller than the symbol size.

FeO has been previously studied with DMC.[1] Therefore, we have only performed selected DMC calculation of FeO to test our Fe-PP. We performed DMC calculation to study fixed-node errors and to evaluate the cohesive energy. These calculations were performed for the experimental structure[2] of FeO (with type-II antiferromagnetic ordering, space group $R\bar{3}m$). Calculations were performed with a 2×2×1 (16 atoms) supercell and twist boundary conditions on a 6×6×6 grid.

The DMC energies of the Fe atom and FeO evaluated with single-particle orbitals generated with LDA+U are shown in Figure S1. Using orbitals from U = 3 eV lowers the

DMC energy of the Fe atom by over 0.3 eV (relative to the DMC energy with orbitals from LDA). For FeO, the larger gain in energy (over 0.8 eV/f.u.) is reached with U = 5 eV or higher. We note that for U-values below 5 eV, LDA+U yields the antiferromagnetic phase of FeO as conducting. The insulating phase of FeO is stable only with U above 5 eV, where a gap of 1.5 eV opens within LDA+U. For U = 7, the band gap from LDA+U is 2.1 eV, which is close to the experimental value of 2.4 eV.

The cohesive energy of FeO was calculated with various time-steps and extrapolated to time-step 0 (Figure S1). Two-body finite size (FS) errors were corrected using the method of Ref. 3 (KZK). For the 16-atom supercell, the calculated two-body FS error is 1.22 eV per formula unit (f.u.). A similar value (1.25 eV/f.u.) was previously reported[4] for the 16-atom supercell of FeO. As shown in Ref. 4, there are residual errors after correcting for the two-body FS. Instead of repeating calculations for FeO with larger supercells to estimate these residual errors, we used the value reported in Ref. 4 (i.e., -0.33 eV/f.u. for 16-atom supercell).

The cohesive energy of FeO evaluated with DMC and single-particle orbitals generated with LDA+U (U = 5 eV) is 9.82(1) eV. Our DMC cohesive energy is very close to the experimental value[5] (9.71 eV). As mentioned above, DMC has been previously employed to study FeO.[1] Those DMC calculations were performed within the locality approximation[6] and with single-particle orbitals generated with hybrid functionals. Instead, we have used the T-moves approach and orbitals generated with LDA+U. Despite these technical differences, the two DMC calculations yield a similar cohesive energy for FeO (the result of Ref. 1 is 9.66(4) eV).

## II. SRFEO₃ AND LAFEO₃ – FIXED-NODE ERRORS

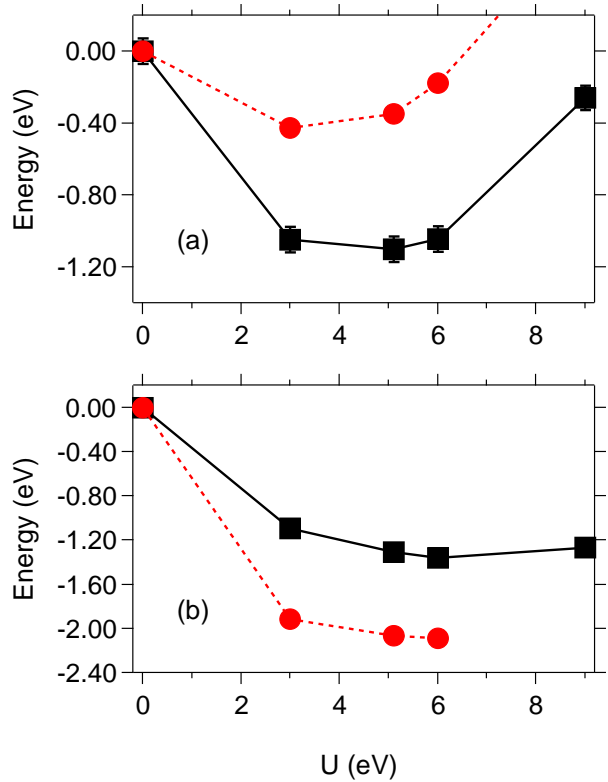

**Figure S2**. DMC energy evaluated with single-particle orbitals generated with LDA+U. Results are included for (a) SrFeO$_3$ and (b) LaFeO$_3$ with (circles) and without (squares) a neutral oxygen vacancy. Energies are given for a $\sqrt{2}\times\sqrt{2}\times2$ (20 atoms) supercell and relative to the DMC energy evaluated with orbitals from LDA. The statistical uncertainty in DMC is smaller than the symbol size in most cases.

We explored fixed-node errors in SrFeO$_3$ and LaFeO$_3$ by generating single-particle orbitals with LDA+U. The DMC energies as a function of U are shown in Figure S2 for SrFeO$_3$ and LaFeO$_3$ with and without a neutral oxygen vacancy. The DMC energy calculated using orbitals generated with U = 3 – 5 and 3 – 6 eV have the lowest values for SrFeO$_3$ and LaFeO$_3$, respectively. Similar values of U also yield the lowest energy for the defected materials. For SrFeO$_3$ and LaFeO$_3$, further DMC calculations were performed with orbitals generated with LDA+U with U values of 3.0 and 6.0, respectively.

## III. SRFEO₃ AND LAFEO₃ – TIME-STEP AND FINITE-SIZE ERRORS

### A. Equation of state

To study time-step errors in SrFeO$_3$ and LaFeO$_3$, DMC calculations were performed at three-time-steps (0.02, 0.01 and 0.005 Ha$^{-1}$). We used a $\sqrt{2}\times\sqrt{2}\times2$ (20 atoms) supercell with experimental lattice constants and twist boundary conditions on a 6×6×4 grid. FS errors (e.g., two-body FS errors),[7] were explored performing an additional DMC calculations with a 2×2×2 (40 atoms) supercell and twist boundary conditions on a 4×4×4 grid. Results of our calculations are shown in upper and central panels of Figure S3 and Figure S4 for SrFeO$_3$ and LaFeO$_3$, respectively. The time-step error in the cohesive of SrFeO$_3$ is 0.21(3) eV (between time-step of 0.01 Ha$^{-1}$ and time-step 0). For LaFeO$_3$, this time-step error is much smaller, only 0.02(3) eV. Employing the method of Ref. 3 (KZK) or the model periodic Coulomb (MPC) interaction[7,8] (corrected for kinetic contributions[7,8]), FS errors can be significantly reduced. The calculated residual FS errors for the 20-atom supercells are 0.63(5) and 0.47(4) eV for SrFeO$_3$ and LaFeO$_3$, respectively. These corrections were applied to the DMC energies of SrFeO$_3$ and LaFeO$_3$ at different volumes (lower panels of Figure S3 and Figure S4. Variation of the residual FS errors due to change in volume was not consider in the present work.

### B. Formation energy of the oxygen vacancy

Time-step errors were study to calculate the formation energy of the oxygen vacancy in SrFeO$_3$ and LaFeO$_3$. These calculations were also performed at three-time-steps (0.02, 0.01 and 0.005 Ha$^{-1}$) with a $\sqrt{2}\times\sqrt{2}\times2$ (20 atoms) supercell on a 6×6×4 grid. Two-body FS errors were calculated employing the KZK and MPC methods for DMC calculations with a 2×2×2 (40 atoms) supercell and twist boundary conditions on a 4×4×4 grid. Results of our calculations are shown in Figure S5.

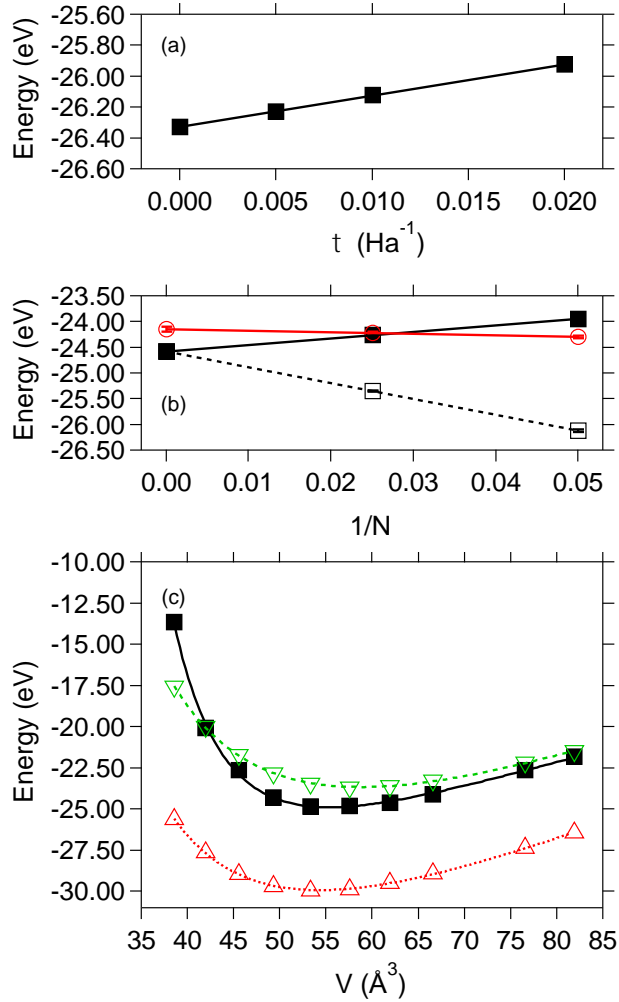

**Figure S3**. (a) DMC energy as a function of imaginary time-step of SrFeO$_3$; extrapolated energies are included at 0 time-step. (b) DMC energy (empty squares), and DMC energy corrected for 2-body FS errors with the KZK method[3] (full squares) and MPC[7,8] (empty circles) as a function of supercell size (cell of 1/40, and 1/20); extrapolated energies are included at $1/N = 0$. (c) DMC (squares), LDA+U (U = 3 eV) (triangles), and PW91+U (inverted triangles) energies versus volume per formula unit of SrFeO$_3$ together with a fitted equation of state (Murnaghan). Note that in (b), the DMC energy has been corrected for time-step error estimated in (a). Similarly, the DMC in (c) has been corrected for time-step error as well as the residual FS errors estimated in (b). Energies are per formula unit and relative to the Sr, Fe and O atoms. The statistical uncertainty in DMC is smaller than the symbol size.

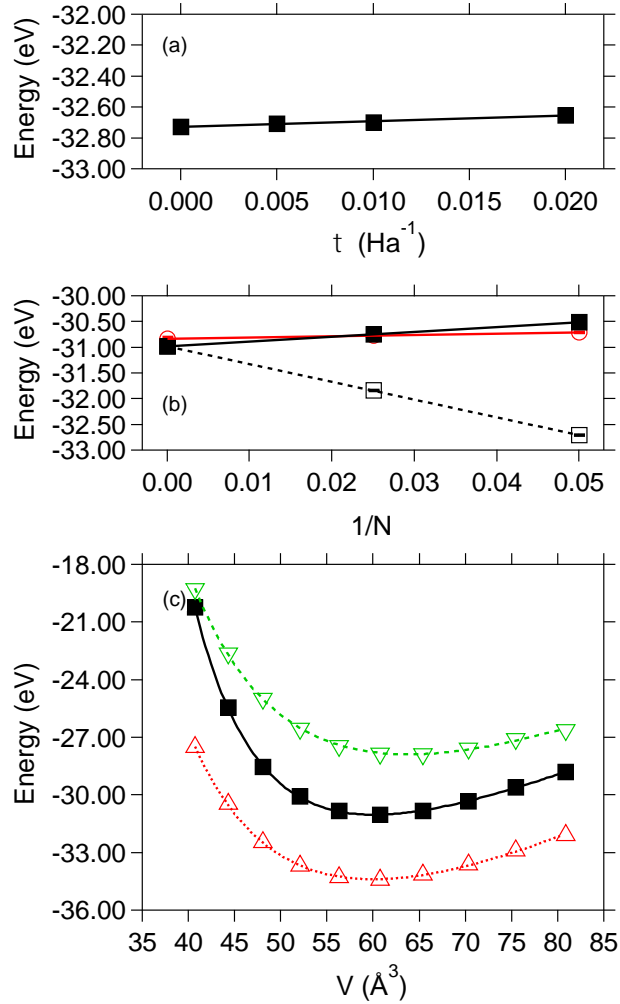

**Figure S4**. (a) DMC energy as a function of imaginary time-step of LaFeO$_3$ (orthorhombic structure with G-type antiferromagnetic ordering); extrapolated energies are included at 0 time-step. (b) DMC energy (empty squares), and DMC energy corrected for 2-body FS errors with the KZK method[3] (full squares) and MPC[7,8] (empty circles) as a function of supercell size (cell of 1/40, and 1/20); extrapolated energies are included at $1/N = 0$. (c) DMC (squares), LDA+U (U = 6 eV) (triangles), and PW91+U (inverted triangles) energies versus volume per formula unit of LaFeO$_3$ together with fitted equations of state (Murnaghan). Note that in (b), the DMC energy has been corrected for time-step error estimated in (a). Similarly, the DMC in (c) has been corrected for time-step error as well as the residual FS errors estimated in (b). Energies are per formula unit and relative to the La, Fe and O atoms. The statistical uncertainty in DMC is smaller than the symbol size.

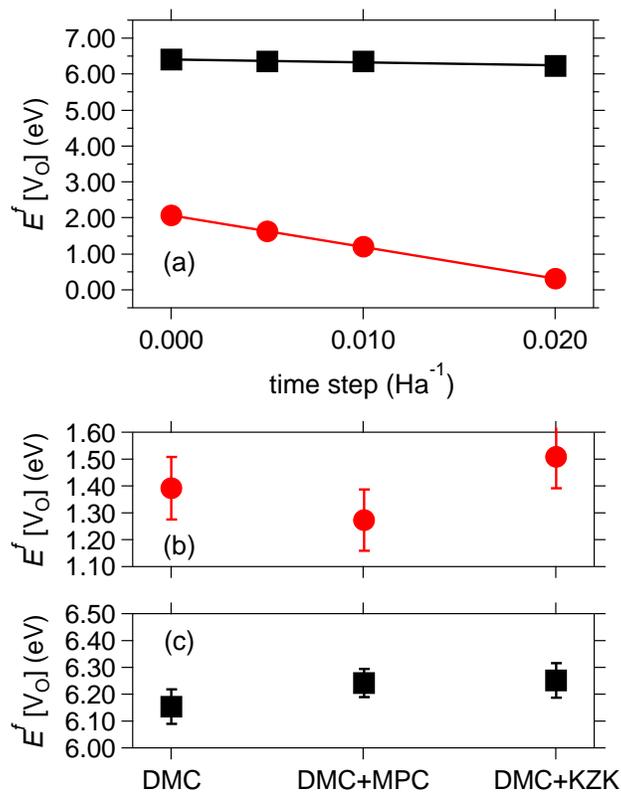

**Figure S5.** (a) DMC oxygen vacancy formation energy in (circles) $SrFeO_3$ and (squares) $LaFeO_3$ as a function of imaginary time-step; extrapolated energies are included at time-step 0. Oxygen vacancy formation energy in (b) $SrFeO_3$ and (c) $LaFeO_3$ evaluated with DMC and DMC corrected for 2-body FS errors with the MPC[7,8] (DMC+MPC) and KZK[3] (DMC+KZK) methods. In (a), the statistical uncertainty in DMC is smaller than the symbol size.

# IV. TABULATE VALUES FOR THE OXYGEN VACANCY FORMATION ENERGIES IN SRFEO$_3$ AND LAFEO$_3$

**Table S1.** Formation energy of the oxygen vacancy in SrFeO$_3$ and LaFeO$_3$ calculated with DMC and DFT approximations. The statistical uncertainty is provided in parentheses.

| Method   | SrFeO$_3$ | LaFeO$_3$ |
|----------|-----------|-----------|
| DMC      | 1.3(1)    | 6.24(5)   |
| LDA      | 1.95      | 4.68      |
| LDA+U(3) | 1.66      | 5.11      |
| LDA+U(5) | 0.97      | 5.15      |
| GGA      | 1.51      | 4.15      |
| GGA+U(3) | 0.92      | 4.38      |
| GGA+U(5) | 0.29      | 4.33      |
| GGA+U(7) | 0.10      | 4.22      |
| HSE06    | 0.17      | 4.56      |